\documentclass[aps,prb,preprint,superscriptaddress,groupeaddress]{revtex4}
\usepackage{graphicx}
\usepackage{amsmath}
\usepackage{graphicx}
\usepackage[Symbol]{upgreek}

\begin{document}

\title{Study of 0-$\pi$ phase transition in hybrid superconductor-InSb nanowire quantum dot devices}

    \author {S. Li}
    \affiliation{Key Laboratory for the Physics and Chemistry of Nanodevices and Department of Electronics, Peking University, Beijing 100871, China}
    \author {N. Kang}
    \email[Corresponding author: ]{nkang@pku.edu.cn}
    \affiliation{Key Laboratory for the Physics and Chemistry of Nanodevices and Department of Electronics, Peking University, Beijing 100871, China}
    \author {P. Caroff $^{a)}$\footnotemark[0]\footnotetext[0]{$^{a)}$Present address: Department of Electronic Materials Engineering, Research School of Physics and Engineering, The Australian National University, Canberra, ACT 0200, Australia}}
    \affiliation{I.E.M.N., UMR CNRS 8520, Avenue Poincar\'e, BP 60069, F-59652 Villeneuve d'Ascq, France}
    \author {H. Q. Xu}
    \email[Corresponding author: ]{hqxu@pku.edu.cn}
    \affiliation{Key Laboratory for the Physics and Chemistry of Nanodevices and Department of Electronics, Peking University, Beijing 100871, China}
    \affiliation{Division of Solid State Physics, Lund University, Box 118, S-221 00 Lund, Sweden}
    \date{\today}

\begin{abstract}

Hybrid superconductor-semiconducting nanowire devices provide an ideal platform to investigating novel intragap bound states, such as the Andreev
bound states (ABSs), Yu-Shiba-Rusinov (YSR) states, and the Majorana bound states.
The competition between Kondo correlations and superconductivity in Josephson quantum dot (QD) devices results in two different ground states and the occurrence of a 0-$\pi$ quantum phase transition. Here we report on transport measurements on hybrid superconductor-InSb nanowire QD devices with different device geometries.
We demonstrate a realization of continuous gate-tunable ABSs with both 0-type levels and $\pi$-type levels. This allow us to manipulate the transition between 0 and $\pi$ junction and explore charge transport and spectrum in the vicinity of the quantum phase transition regime.
Furthermore, we find a coexistence of 0-type
ABS and $\pi$-type ABS in the same charge state. By measuring temperature
and magnetic field evolution of the ABSs, the different natures of
the two sets of ABSs are verified, being consistent
with the scenario of phase transition between the singlet
and doublet ground state.
Our study provides insights into Andreev transport properties of hybrid superconductor-QD devices
and sheds light on the crossover behavior of the subgap spectrum in the vicinity of 0-$\pi$ transition.
\end{abstract}

\maketitle

\begin{center}\noindent{\bf I. INTRODUCTION}\end{center}

In hybrid structures where a superconductor (S) is connected to a mesoscopic
normal conductor (N), an electron in the normal region is converted
to a Cooper pair into the superconductor with the reflection of a
hole. This well established mechanism, known as the Andreev reflections,
plays a central role in the proximity effect in which the electron
pairing potential in S can penetrate into N region. During the past
years, based on the idea of introducing superconducting order into
low-dimensional semiconductors via the proximity effect, superconducting
leads coupled semiconductor quantum dots (QDs) have been extensively
investigated for novel quantum phenomena and new device concepts\cite{SchoNN},
such as tunable Josephson junctions\cite{Doh,Jarillo,Xiang,Cleuziou,Nilsson} and Cooper-pair
splitters\cite{Hofstetter,StrunkCPS,MotyCPS,DeaconCPS}, and such
hybrid S-QD structures provide an ideal platform to study basic physical
issues including the formation of Andreev Bound States\cite{Bauer,Deacon,Pillet,PilletB,Simon,Ingap},
the interplay between the Kondo effect and the proximity induced superconductivity\cite{Rozhkov, LevyPi, Egger, Belzig, Choi, Buitelaar, Tarucha07, Eichler07, Jespersen07, silvanoPRL, Kim, Kanai10},
as well as the search for Majorana fermions in solid state\cite{Lutchyn,Oreg}.
Andreev Bound States (ABS), originating from the superposition of coherent
Andreev reflection processes, have been attracting growing interests
owing to its fundamental importance in mesoscopic superconductor related
systems, and have been observed in both S-QD-S \cite{Pillet,PilletB}and
S-QD-N\cite{Deacon,Willy,silvanoNN} systems in the tunneling regime.
In superconducting leads coupled QDs, the spectrum of ABSs can be
dramatically influenced by Coulomb interaction and spin-related many-body Kondo correlation.
Depending on the ratio of the corresponding
energy scales $k_{B}T_{K}/\Delta$ where $T_{K}$ is the Kondo temperature and $\Delta$ is the superconducting gap, a 0-$\pi$ Josephson junction transition
has been predicted theoretically\cite{Rozhkov,Ingap,Simon} and observed
experimentally by either the sign change of the Josephson supercurrent\cite{Reversal, Jorgensen, Eichler09, Squid}
or the crossing behavior of the ABSs\cite{Deacon,Kim,silvanoNN}.
When $k_{B}T_{K}/\Delta\ll1$,
the Kondo screening is suppressed due to the lack of quasiparticle
DOS around the Fermi level, hence the junction is in a $\pi$ state
with a crossing of the two Andreev levels. For $k_{B}T_{K}/\Delta\gg1$,
a Kondo singlet forms by breaking cooper pairs at the Fermi level, thus the junction is in a 0 state and the two Andreev levels never cross.
Since these two magnetic states - a doublet
for the $\pi$ state and a singlet for the 0 state - are
two different ground states (GSs) of the QD with different spins, a quantum phase transition is expected to be induced by a change of parameters in the QD device, such as the chemical potential and the charging energy.

InSb NWs own a giant g factor\cite{Giant}
and strong spin-orbit interaction strength\cite{Giant,NadjSOI, Fan}. Hybrid superconductor-InSb NW devices are expected to exhibit rich physics of subgap states
due to the influence of strong spin-orbit coupling
and Zeeman splitting in the presence of magnetic fields. Recently,
zero-energy states have been observed in hybrid superconductor-InSb NW
devices and been interpreted as Majorana Bound States\cite{Mourik,Mingtang,Mingtang1},
though under certain circumstances ABSs could behave as imitations of
Majorana zero-energy states, yet having no relations with topological superconducting
phases. Therefore to clarify the mechanism of these subgap bound
states in such systems, a systematic study of the ABSs in hybrid superconductor-InSb NWQD devices is of fundamental importance, and is still lacking.

In this work we report an extensive study on the low-temperature transport measurements
of hybrid superconductor-InSb NWQD devices with different device
geometries, i.e. the InSb nanowire QD(NWQD)-SQUID device and the S-NWQD-S
device. In the NWQD-SQUID device we observe continuous gate-tuned ABSs with both
0-type levels and $\pi$-type levels
and study the evolution of the ABSs as a function of temperature,
showing a tunneling transport mechanism assisted by the thermally
populated ABSs.
In the S-NWQD-S device, we demonstrate that two sets of 0-type
and $\pi$-type ABSs overlap at the same Kondo region, though a doublet
ground state is favored given the estimated Kondo temperature. We propose a possible
scenario for the simultaneous emergence of the 0-type and $\pi$-type
ABSs and this scenario can be supported by the temperature and magnetic
field evolution of the ABSs.

\begin{center}\noindent{\bf II. EXPERIMENTAL DETAILS}\end{center}
The hybrid superconductor-InSb nanowire QD devices are fabricated
from individual InSb segments of InAs/InSb heterostructure NWs grown
on an InP substrate by MBE method\cite{XuNanotech2012, Thelander2012}, with typical diameters ranging
from 60 to 120 nm. After growth, the NWs are transferred onto a degenerately
doped, n-type Si substrate (used as a global backgate), covered by
a 110-nm thick thermal oxide. The superconducting electrodes (5 nm/100
nm Ti/Al bilayer) are fabricated using standard e-beam lithography
and e-beam evaporation procedure. Before the metal deposition, the
samples are etched in a diluted $(NH_{4})_{2}S_{x}$ solution to remove
the native oxide layer. The Ti layer is used as an adhesion layer
between the superconducting Al layer and the nanowire. All measurements
are performed in a $^{3}$He/$^{4}$He dilution refrigerator with
a base temperature of $T\sim10$ mK. In order to minimize the electronic noise, we used a series of $\pi$-filters, copper-powder filter and RC filters thermalizing at different temperature stages
The magnetic field is perpendicularly
applied to the sample plane.

\begin{center}
\noindent{\bf III. ANDREEV BOUND STATES IN NWQD-SQUID DEVICES}
\end{center}

\begin{center}
\noindent
\textbf{A. Gate-tuned different types of ABSs in NWQD-SQUID devices.}
\end{center}


Fig. 1(a) shows a scanning electron microscope (SEM) picture of a typical InSb NWQD-SQUID device and a schematic of the measurement circuit.
This geometry is commonly used to improve the energy resolution for probing subgap resonances of the QDs due to the sharp change in BCS density of states at the superconducting gap edges\cite{Pillet, PilletB, Willy, Kumar, silvanoPRL}.
In our case, as can be deduced from the Coulomb diamond (see section I of supplementary material),
the central electrode is weakly coupled to the nanowire and plays the role of probing the DOS of levels in the nanowire. The top panel of Fig. 1(d) shows the differential conductance $dI/dV$ in the superconducting state as a function of bias voltage $V_{sd}$ and backgate voltage $V_{g}$, acquired
at base temperature and zero magnetic field. We can clearly see
two main features over the entire $V_{g}$ range: a pair of main resonances at higher bias voltages
with high conductance ($\circ$) and a pair of weak resonances with
low conductance ($\diamond$) lying parallel to the main resonance.
We attribute the main resonances to the alignment of the ABS levels
at $\pm\varepsilon_{a}$ with the singularity in the BCS density of
states of quasiparticles, i.e. at bias voltages: $eV_{sd}=\pm(\varepsilon_{a}+\Delta)$
(see Fig. 1(b)). The weak resonances can be interpreted as a replica of the
main ABSs when the residual quasiparticle DOS of the probe at the
Fermi level is aligned with the ABS levels: $eV_{sd}=\pm\varepsilon_{a}$ \cite{Kumar}.

In particular, we can readily identify distinct gate voltage evolution of the ABSs resonances
in charge states of different parity determined by alternate size of the Coulomb diamond patterns.
In regions with even electron occupation, the main ABSs resonances appear to overlap
with the elastic quasiparticle cotunneling at $eV_{sd}=\pm2\Delta\approx\pm300 \ \upmu$eV,
showing little dependence on gate voltage. This denotes that the ABSs
are pushed towards the superconductor gap edge in the even states.
Fig. 1(e) shows a $dI/dV(V_{sd})$ trace taken at the center of the even valley indicated by the orange dashed line in Fig. 1(d).
The positions of peaks in $dI/dV$ indicate the superconducting gap, $\Delta=1.76k_{B}T_{c}\approx150 \ \upmu$eV,
which correspond with a critical temperature of $T_{c}\approx1$ K.
In contrast, in the odd charge states, the shape of ABSs peaks is evidently dependent
on the gate voltage and shows either crossing (blue arrow) or noncrossing (green
arrow) behavior within different odd-occupied Coulomb valleys.
Fig. 1(f) shows a $dI/dV(V_{sd})$ trace taken at the center of the odd valley indicated by the light-blue dashed line in Fig. 1(d).
A differential conductance peak appears symmetrically at finite bias voltages inside the superconducting gap, $|V|<2\Delta$, and a negative differential conductance (NDC) dip develops at bias voltages near $\pm2\Delta$.
The presence of NDC can be explained by the asymmetric coupling between the QD and two superconducting leads, playing the role of probing ABSs \cite{Andersen, Kim}.
The characteristic subgap structures have been reported in hybrid superconductor-QD devices in the odd occupation regions\cite{Deacon, Grove}, which can be ascribed to spin-induced subgap states forming in the QD energy spectrum.

As previous works have
studied\cite{Rozhkov, Egger, Belzig, Choi, Buitelaar, Tarucha07, Eichler07, Jespersen07, silvanoPRL, Kim}, these different gate-dependent behaviors
of ABSs (crossing or noncrossing) in the odd charge states can be understood
as a competition between different energy scales, namely, Coulomb interaction $U$, QD energy level $\xi_{d}$, and hybridization $\Gamma$ to superconducting leads.
The phase diagram is displayed in Fig. 1(c).
In the $k_{B}T_{K}/\Delta\gg1$ limit, the ground state of the system
is always a singlet. Thus the system stays in 0 state and the
ABS levels demonstrate a noncrossing behavior.
In the opposite limit of $k_{B}T_{K}/\Delta\ll1$, the system alters its ground state from
a singlet to a magnetic doublet as the ABS levels cross at zero energy.
By sweeping gate voltages, the strength of the coupling between the leads and the QD can be tuned, resulting in different Kondo temperature $T_{K}$.
As approaching the singlet-doublet phase boundaries, the position of ABS energy $\varepsilon_{a}$ moves towards zero, and a 0-$\pi$ quantum phase transition is expected to occur.
The bottom panel of Fig. 1(d) displays gate voltage dependence of the zero-bias conductance.
It can be readily seen that the conductance is enhanced at gate positions where the ABS levels cross zero energy and the system undergoes a parity-changing quantum phase transition.
The most surprising result is the enhanced zero-bias conductance of large amplitude in the Kondo valley at -0.59 V $<V_{g}<$ -0.53 V.
In Fig. 1(g) we show conductance line cut through the center of Kondo valley indicated by the green dashed line in Fig. 1(d).
It clearly displays a sharp zero-bias conductance peak and two subgap peaks at $V_{sd}\sim\pm\Delta/e$, corresponding to $\varepsilon_{a}\sim 0$. To clarify the nature of this zero-bias conductance peak, we perform current bias measurements and show the $I-V$ curve in the inset of Fig. 1(g). A typical $I-V$ shape of an overdamped Josephson junction comprising a supercurrent branch and a dissipative branch can be recognized, with a switching current $I_{SW}\sim 1$ nA which is consistent with typical magnitude in similar nanowire Josephson junctions\cite{SenInSbJJ}. Hence, the zero-bias conductance peak can be attributed to the manifestation of a supercurrent and this is also confirmed by data from a second similar device (see Fig. S2(c) in Supplementary Materials). We argue that such supercurrent peak is further enhanced by the Kondo effect near the 0-$\pi$ phase boundary\cite{GroveSC}.
Similar behavior of Kondo-enhanced Andreev transport at finite bias has also been reported in hybrid superconductor-QD devices\cite{Eichler07, Jespersen07}, which can be contributed to a logarithmic enhancement of transparency, $1/\ln^{2}(\Delta/k_{B}T_{K})$, by a poor man's analysis.
We speculate that a singularity of Kondo-enhanced transparency at $k_{B}T_{K}\sim \Delta$ can be responsible for the enhanced zero-bias conductance in a crossover region from a singlet to a doublet ground state.

\begin{center}
\noindent
\textbf{B. Thermally excited Andreev spectra.}
\end{center}

Having characterized the gate-tunable 0-$\pi$ transition of the Josephson junction in a InSb NWQD device, we now focus on the thermal effect on the two types of ABSs.
Figs. 2(a) and 2(c) show the differential
conductance $dI/dV$ in two types of Kondo valleys as a function of bias
voltage $V_{sd}$ and backgate voltage $V_{g}$ measured at base temperature and zero magnetic field.
As mentioned above, depending on the respective ratio of
$k_{B}T_{K}/\Delta$, the main ABSs resonances behave as a crossing
in Fig. 2(a), indicating the charge state is in a $\pi$ state in
the middle of this region, while in Fig. 2(c) the noncrossing ABSs resonances denote
a 0 state of that region. When the temperature is elevated gradually
up to 700 mK, several prominent features of the ABSs resonances emerge
in both $\pi$ state and 0 state regions.
First we look at the $\pi$ state
case (Fig. 2(b)). As the temperature is increased to 700 mK, the inner
weak replicas -- stem from the the alignment of residual density of
states of quasiparticles in the probe at the Fermi level with the ABS
levels - almost disappear. This could be explained by the thermal
smearing of the quasiparticle DOS at the Fermi level. Moreover, the
$\pi$-type ABSs in the center of the Coulomb valley extend out of
the $\pi$ state region while the 0-type ABSs outside of the crossing
points penetrate into the $\pi$ state region, these extending structures
form continuous resonances which are parallel to the original ABSs
resonances at opposite $V_{sd}$ at base temperature. For the 0 state
case in Figs. 2(c) and 2(d), beside the main ABSs peaks, there are no inner
structures at the base temperature. However, a new pair of weak resonances
develop as the temperature is increased to 700 mK and similarly to
the $\pi$ state case in Fig. 2(b), the newly developed resonances are
parallel to the original ABSs resonances at opposite $V_{sd}$.
A more detailed study of the ABSs spectroscopy at a series of elevated temperatures is given in section III of supplementary material.
We note that both the main ABSs resonances and their replicas shift towards
lower energies in the same way due to the shrink of $\Delta$ as the temperature is
increased, suggesting the same origin of these subgap features.

The evolution of the observed ABSs spectroscopy on temperature can be understood in term of thermally excited quasiparticle transport in the subgap region\cite{Kumar}.
As the temperature is elevated, quasiparticles can be thermally excited from the lower electron-like
ABS to the upper hole-like ABS. As a consequence, charge transport through the thermally populated
ABS level could be expected to generate additional resonant peaks in differential
conductance at finite bias when the singularity of the BCS density of states of the
probe is aligned with the thermally populated ABS. Hence these resonances
are supposed to occur at $V_{sd}(V_{g})=\pm[\Delta-\varepsilon_{a}(V_{g})]$.
Given the condition for the original ABS resonances to occur, i.e. $V_{sd}(V_{g})=\mp[\Delta+\varepsilon_{a}(V_{g})]$,
the position of the thermally developed resonance and that of the
original ABS resonance at finite bias corresponding to the same ABS level are spontaneously
parallel with a constant spacing of 2$\Delta$. Therefore, these thermally-induced subgap features in the Andreev spectrum
can also be viewed as replicas of the original
ABS resonances.
We also note that the width of the newly developed resonances exhibits a clear level broadening with increasing temperature, being consistent with the thermally populated ABS model.
Compared with previous study, e.g. ref [37], we have observed the thermally developed ABS resonances in both $\pi$-type ABS region and 0-type ABS region, thus constituting a complete experimental picture of the thermal population of different types of ABSs.

\begin{center}
\noindent{\bf IV. ANDREEV BOUND STATES IN THE S-NWQD-S DEVICE}
\end{center}

\begin{center}
\noindent
\textbf{A. Coexistence of 0 and $\pi$-type ABSs in the S-NWQD-S device.}
\end{center}

We now consider another case of a S-NWQD-S device and focus on the
formation (and the underlying physical mechanism) of the ABSs with
unusual appearances. We first characterize the device with the leads in the normal state. Fig. 3(a) presents differential conductance $dI/dV$
as a function of $V_{sd}$ and $V_{g}$ measured at the base temperature with a small magnetic field applied ($B=50$ mT).
For the right region
at $V_{g}\gtrsim $ -1.2 V, the transport is in a relatively open regime, where
the average differential conductance is high and exhibits a series of broad resonances at low bias voltages.
As the gate voltage becomes more negative, the differential
conductance becomes lower with the emergence of a characteristic Coulomb diamond
in the 2D plot, denoting the formation of a well defined QD.
The charging energy can be estimated from the size of diamond as $U\sim2.3$ meV.
Looking more closely at the diamond region, two narrow ridges of high conductance around zero bias, indicated by the black arrows, can be clearly seen in the Coulomb valley (see Fig. S4(b) in supplementary material), suggesting the emergence of a Kondo resonance.
By measuring the temperature dependence and magnetic field evolution
of the $dI/dV$ in the middle of the odd region (see Fig. S4(c) in supplementary material), we further verify that
this split structure is a Kondo feature split by the Zeeman energy resulting
from the large g-factor in InSb. From the magnetic field dependence
of the Kondo splitting, we extract an effective \emph{g}-factor of this Kondo
valley: $|g^{\ast}|\sim35$ (see Fig. S4(d) in supplementary material).
From the Lorentzian fit of the zero-bias conductance
in the middle of the Kondo valley as a function of the magnetic field,
the zero-bias conductance $G_{max}$ of Kondo ridge at $B=0$ is estimated
to be $0.19\ e^{2}/h$, and the asymmetry ratio $\gamma=\Gamma_{L}/\Gamma_{R}\sim40$
is obtained from the relation $G_{max}=(2e^{2}/h)4\Gamma_{L}\Gamma_{R}/(\Gamma_{L}+\Gamma_{R})^{2}.$\cite{Gmax}
This high asymmetry ratio indicates that our device can be viewed as a tunneling probe model in which
the ABSs are generated mainly between the QD and the strongly coupled lead, and are probed by the other
weakly coupled lead\cite{Deacon,Kim,Tarucha07,Eichler07}.


Next we turn to look at the superconducting state spectrum of the
device. Differential conductance $dI/dV$ as a function of $V_{sd}$
and $V_{g}$ of the Kondo region measured at $B=0$ and the base temperature is
displayed in Fig. 3(b). While the transport at the gate region
away from the left resonant point ($V_{g}< -1.6$ V) is dominated by multiple Andreev reflections,
with a quasiparticle cotunneling peak at $|eV_{sd}|=2\Delta$ and
a weak first-order multiple Andreev reflection peak at $|eV_{sd}|=\Delta$,
the Kondo valley exhibits rich subgap structures. Beside the weak
first-order multiple Andreev reflection at $|eV_{sd}|=\Delta$, we
concentrate on the main ABS resonance peaks which vary in the finite bias
range from $|eV_{sd}|=2\Delta$ to $|eV_{sd}|=\Delta$ as a function
of the gate voltage. The most noticeable feature is that there are
two different types of ABSs peaks of opposite curvature in dispersion
of $V_{g}$ coexisting in this Kondo valley region, namely the $\pi$-type
crossed ABSs at outer positions(green dashed lines) and the 0-type noncrossed ABSs at inner positions (black dashed lines). This is more clearly characterized by two pairs of
subgap $dI/dV$ peaks in Fig. 3(c), showing the line cut of $dI/dV$
in the middle of the Kondo valley in Fig. 3(b).
In addition, these ABSs
resonances appear to extend into the open regime region next to the Kondo
valley, forming a set of weak ABSs resonances in the open region.
As we have mentioned, the dispersion of two types of ABSs in the Kondo states
is dominated by the system's ground state - singlet or doublet - depending on the ratio $k_{B}T_{K}/\Delta$.
Taking total coupling strength $\Gamma$
as determined from the Coulomb resonance peaks of \emph{N} state, the Kondo temperature at the middle of
this valley can be estimated from the Bethe-Ansatz model\cite{Eichler09, Bethe} as: $T_{K}\sim460$ mK (see section IV of supplementary material),
which is much smaller than $\Delta/k_{B}$.
From the magnetic field evolution of $dI/dV(V_{sd})$, a splitting of the Kondo resonance is observed already
at a minimum field $B\sim25$ mT, corresponding to a Zeeman energy
of $E_{Z}\sim49\ \upmu$eV, providing further evidence of the small value of $T_{K}$.
Therefore, this odd charge state
is expected in a $\pi$ state, and one would expect only a pair of $\pi$-type
ABSs, being inconsistent with experimental observation.
To understand the observed coexistence of the two types to ABSs, we propose a simple model, in which the transport processes in our device involve two parallel conduction channels with different coupling strength to the superconducting leads, which is illustrated in the inset of Fig. 3(c).
Here, one channel is weakly coupled to the electrodes and sustains the weak Kondo correlation, leading to the formation of $\pi$-type ABS.
The other channel, which is strongly coupled to the electrodes, accounts for the
emergence of the 0-type ABS. We now discuss the detailed mechanism about the two channel model in the following paragraph.

In fact, looking more closely at the normal state conductance plot in Fig. 3(a), we can see
some broad resonances with crossed diagonal shapes at low bias on the right side of the Kondo charge state.
For better clarity, we take a line cut in this region (grey dashed line) and plot the $dI/dV(V_{sd})$ curve in Fig. 3(d). A set of quasi-periodic broad resonances at low bias could be clearly recognized.
We argue that these broad resonances is the manifestation of the Fabry-P\'erot interference in a ballistic nanowire.
Fabry-P\'erot interference patterns have been observed in InSb nanowire Josephson junctions in our previous work\cite{SenInSbJJ} and a mean free path $l_{e}\sim 80$ nm of our InSb nanowires has been extracted.
Given here the channel length $L\simeq 65$ nm$<l_{e}$, the NWQD device is supposed to work in the ballistic
regime and the occurrence of the Fabry-P\'erot interference in the open regime is quite reasonable.
Moreover, we can extract the average energy spacing of the Fabry-P\'erot resonances from Fig. 3(d) as
$\Delta E\sim2.5$ meV. Using the expression of energy spacing of the Fabry-P\'erot interference in a 1D subband $\Delta E_{FP} = \hbar^{2}\pi^{2}/2m^{*}L_{c}^{2}$, where $m^{*}$ is the electron effective mass and $L_{c}$ is the cavity length, we calculate an effective cavity length $L_{c}\sim75$ nm, which is quite consistent with the geometric channel length of the device and thus supports our interpretation. Therefore, beside the weakly coupled
quantum dot level, an additional channel with stronger coupling constituted by the Fabry-P\'erot resonant states could also contribute to transport. Since the device shows a quick transition from the Coulomb blockade regime to open transport regime and the Kondo valley is just next to the transitional region (Fig. 3(e)), it is very likely that these two channels with different coupling strengths are involved in the transport simultaneously, which is also supported by the $dI/dV$ curve in Fig. 3(d) where the Fabry-P\'erot resonances at low bias and the sharp tunneling resonance at high bias are observed simultaneously. Hence, the coexistent 0-type and $\pi$-type ABS could be reasonably understood on the basis of such a two-channel model.

\begin{center}
\noindent
\textbf{B. Temperature evolution of the coexistent 0 and $\pi$-type ABSs.}
\end{center}

To better understand the properties of  the coexistent 0 and $\pi$-type ABSs,
we measure the differential conductance $dI/dV(V_{sd})$
along the cut in the middle of the Kondo valley (blue dashed line in Fig.
3(b)) over a wide range of temperature.
The colored $dI/dV$ curves
shown in Fig. 4(a) correspond to different temperatures, ranging from
base temperature ($\sim$10 mK) to the critical temperature
of the Al leads ($\sim$ 1 K), and they are plotted as a 2D
graph in Fig. 4(b) (a temperature evolution of the overall  $dI/dV(V_{sd},V_{g})$ spectra is given in Fig. S5 in supplementary material). From Figs. 4(a) and 4(b), we can see that the differential
conductance $dI/dV$ barely changes in a low temperature range, i.e.
$T\lesssim300$ mK.
However, as the temperature is further elevated, several
features become pronounced. First, the most prominent feature is the
emergence of a zero bias conductance peak (ZBCP), rising up from approximately
400 mK. An ABS-assisted resonant tunneling process can lead to such ZBCP, as depicted in Fig. 4(c).
Specifically, quasiparticles
in the superconducting leads could be thermally excited from the continuum
band below the Fermi level $E_{F}$ to the empty band above $E_{F}$ as temperature increases,
occupying the density of state near the upper gap edge. Meanwhile
the same process occurs between a pair of ABSs, leaving each level
of the ABSs partially occupied. Note that inside the Kondo region
the position of the $\pi$-type ABSs is very close to the superconducting gap edge.
Thus the thermally excited quasiparticles in the superconducting contacts
could tunnel through the quantum dot via both $\pi$-type
ABSs near the two edges of the gap.

Although the scenario presented above is most likely
to account for our observation, we also note that some features of the temperature dependence of the differential conductance in Fig. 4(a)
can't be fully explained by a thermally activated process.
First, we find the width of the ZBCP changes little with increasing temperature, as indicated by the dotted line in Fig. 4(a).
If the observed ZBCP is due to thermally excited quasiparticle transport, then the width of conductance peak should be broadened with increasing temperature, inconsistent with experimental result.
Second, the temperature dependence of the ZBCP does not follow an exponential behavior as expected for a thermally activated process.
In Fig. 4(d) we plot $G_{0}/G_{N}$ as a function of
$ k_{B}T_{K}/\Delta$, where $G_{0}$ is the zero bias conductance and $G_{N}$ is the normal state conductance
at high bias of each curve in Fig. 4(a). The ratio of $G_{0}/G_{N}$ first rises rapidly at small $k_{B}T_{K}/\Delta$, then tends to
saturate as $k_{B}T_{K}/\Delta$ becomes larger corresponding to the increase of temperature. The transition of these two trends occurs
at roughly $k_{B}T_{K}/\Delta \sim 0.3$, which is coincident with theoretical predictions for the 0-$\pi$ transition to occur\cite{Bauer, Belzig}.
In a weak coupling regime,  $k_{B}T_{K}/\Delta \ll 1$, the system is in the $\pi$-state.
As the temperature is raised, $\Delta$ is reduced changing the relative strength of $\Delta$ and $k_{B}T_{K}$, undergoing a crossover from $\pi$ state to 0 state.
In the 0 state, Kondo resonance develops even in the presence of the superconducting gap, and contributes an conductance peak in equilibrium regime.
Therefore, we speculate that it is likely that a temperature induced 0-$\pi$ transition is also
responsible for the emergence of the observed ZBCP.
Another interesting appearance, shown in Fig. 4(b), is that the 0-type ABSs at finite bias slightly shift
to higher energy as the temperature increases above 300 mK.
To understand this behavior, we
look at the schematic phase diagram in the parameter space $(\Delta/\Gamma, U/\Gamma)$ shown in Fig. 4(e), where
$\Delta, \Gamma, U$ represents the superconducting gap, the coupling strength between QD level and superconducting leads, and the charging energy, respectively\cite{Bauer, Squid}.
When temperature is increased in the range of $\sim1$ K, $\Delta$ is suppressed while no noticeable change of $\Gamma, U$ is expected,
resulting in a reduced ratio of $\Delta/\Gamma$, represented by the red dashed line in the phase diagram in Fig. 4(e).
Theoretically the ABSs tend to approach the continuum of the BCS band in the strong coupling
limit $\Delta/\Gamma\ll1$\cite{Bauer, Simon}, hence the shift of 0-type ABSs towards higher energy (gap edge) with increasing temperature can be reasonably understood.

\begin{center}
\noindent
\textbf{C. Magnetic field evolution of the coexistent 0 and $\pi$-type ABSs.}
\end{center}
Finally, we explore the effect of magnetic field on the spectroscopy of ABSs.
Fig. 5(a) displays the magnetic field
evolution of the differential conductance measured along the blue dashed line cut in Fig. 3(b). As the magnetic field is
applied, the energy of both $\pi$-type ABSs and 0-type
ABSs tends to decrease.
Surprisingly the $\pi$-type ABSs seem
to decrease in an unusual fast way and become hardly distinguishable
at $B\sim10$mT -- far below the superconductor critical
field $B_{c}\approx25$ mT. The 0-type ABSs, on the other hand, vary
much more slowly compared with the $\pi$-type ABSs and remain
to be resolved as the magnetic field approaches $B_{c}$. Above the
critical field, the ABSs peaks are replaced by a split Kondo resonance.
In the presence of a magnetic field, two mechanisms that may have impact
on the ABSs should be considered. On the one hand,
the superconducting gap $\Delta$ is suppressed by increasing magnetic
field, which should result in a subsequent suppression of the intragap states and
may also change the relative strength between Kondo correlation and
superconductivity, making it possible for a 0-$\pi$ transition to
happen\cite{silvanoPRL}.
But this effect is not supposed to be obvious
until the magnetic field approaches $B_{c}$ since $\Delta$
changes little at magnetic fields far below $B_{c}$.
On the other hand, the Zeeman effect will alter the energy of different types
of ABSs, depending on the magnetic property of corresponding ground
state, i.e. singlet or doublet. For a relatively small magnetic field,
few tens of mT in our case, one might think of negligible effect of
the Zeeman energy. However, considering the large g factors in InSb
nanowires, the role of the Zeeman effect can not be neglected.

The energy of the $\pi$-type ABSs is expected
to increase in a magnetic field as a result of the Zeeman splitting of its doublet
GS\cite{silvanoNN} (see lower right inset of Fig. 5(b)). But in our
case, the $\pi$-type ABS is quite close to the superconducting gap
edge. Taking the level-repulsion effect\cite{silvanoNN} between the
ABSs and the continuum of quasiparticle states into account, it is
reasonable that we could not observe an increase of the energy of $\pi$-type
ABSs. As already mentioned, the $\pi$-type ABSs are related with the QD level that is weakly coupled to the superconducting contacts. Such weakly coupled level may be localized in the NW segment just in the middle of the junction by the thick barriers, while the strongly coupled level related with the 0-type ABSs could extend in a longer geometric range. Hence, we attribute the rapid shrink of the $\pi$-type ABS to the existence of
magnetic flux focusing from the Meissner effect in the case of a narrow Josephson junction
geometry\cite{Giazotto}.
As for the 0-type ABSs, in a finite magnetic field, they will also follow the suppression of $\Delta$. Besides, a splitting in such ABSs is also expected due to the Zeeman splitting
of doublet ES (upper left inset of Fig. 5(b)). In our data, the 0-type
ABSs show an overall decreasing trend as $B$ increases. Considering
the relatively low magnetic field we apply limited by the critical
field of Al, in such magnetic field range, the shrink of the gap is
much more significant than the Zeeman energy. Thus the level repulsion
effect between the ABS and the continuum of quasiparticle states leads
to the overall decreasing trend of the 0-type ABSs. Nevertheless,
we note that there is a sign of Zeeman splitting in the
0-type ABSs at $B\sim14$ mT indicated by the black arrows in Fig. 5(a). The corresponding
$dI/dV(V_{sd})$ is shown in Fig. 5(b), in which we are able to identify
a split double-peak on both 0-type ABSs peaks. The energy distance
between the double peaks is measured as $\Delta E\sim34\ \upmu$eV. Using
$|g^{*}|\approx35$ estimated from the Zeeman splitting of the Kondo
resonance, we obtain a Zeeman energy $E_{z}=|g^{*}|\mu_{B}B\approx31\ \upmu$eV
at $B=14$ mT, which is consistent with the spacing of the double peaks,
further supporting our interpretation.

\begin{center}
\noindent
\textbf{V. CONCLUSION}
\end{center}

In summary, we demonstrate gate tunable ABSs of different types in
superconductor coupled InSb nanowire QDs with different device structures
by transport measurements. The thermal effect on the two types of ABSs
is extensively explored in the NWQD-SQUID devices and is understood
with a thermally excited quasiparticle model. Two types of ABSs are
observed simultaneously in the same charge state in a S-QD-S device,
and explained in the senario of two conducting levels with different coupling strength to the leads.
The evolution in elevated temperatures and magnetic field
confirms the different natures of these ABSs. Therefore, despite the different device geometries,
the spectroscopy of two types of ABSs in superconductor coupled single InSb QDs
has been unambiguously displayed. Together with the variation of these ABSs,
the 0-$\pi$ transition process driven by different physical parameters - gate voltage (in SQUID device) or temperature (in S-NWQD-S device) - has been revealed in such hybrid systems.
Moreover, our work is the first systematic study of the tunable subgap bound states and 0-$\pi$ transition in InSb nanowire based hybrid superconducting systems which is an important candidate
to host Majorana bound states. These results demonstrate
the key role of Andreev Bound States in the transport in mesoscopic
Josephson junctions and indicate potential prospect in exploring intragap
bound states such as Majorana bound states in such systems.

\begin{center}
\noindent
\textbf{ACKNOWLEDGMENTS}
\end{center}
We thank D. X. Fan, G. Y. Huang and L. Lu for helpful discussions.
This work was financially supported by the National Basic Research Program of the Ministry of Science and Technology of China (Nos. 2012CB932703 and 2012CB932700), and by the National Natural Science Foundation of China (Nos. 11374019, 91221202, 91421303, and 61321001). H. Q. Xu acknowledges also financial support from the Swedish Research Council (VR).

\newpage

\begin{center}\noindent{\bf References}\end{center}


\newpage

\begin{figure}

\centering{}\includegraphics[width=0.73\textwidth]{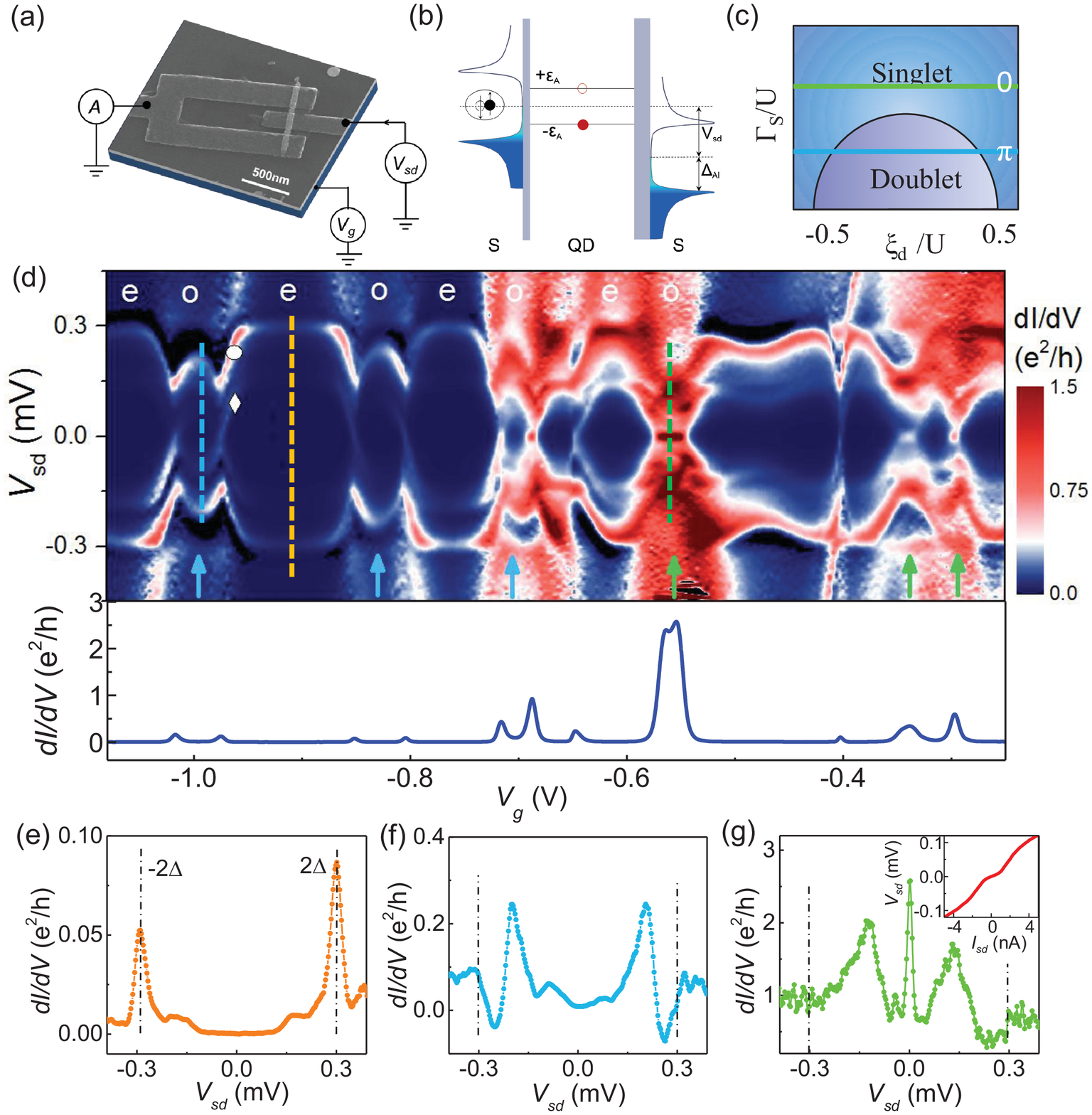} \caption{Gate tuned ABSs in the NWQD-SQUID device. (a) SEM image of an InSb NWQD-SQUID device and schematic of
the measurement configuration. The diameter of the nanowire is $\sim$ 80 nm and
the channel length of the nanowire is designed to be around 100 nm.\textbf{
}(b) Schematic of tunneling processes in the NWQD-SQUID system. The alignment of the singularity
of the hole(electron)-like quasiparticle DOS of the probe with the
electron(hole) type Andreev bound states results in the main resonance
peaks in the $dI/dV$ spectra. (c) Phase diagram of superconductor coupled
QD system. $\xi_{d}$ represents the relative position of the chemical potential of the
QD, $U$ is the charging energy of the QD, and $\Gamma_{s}$ is the
coupling strength of the QD with superconducting leads. (d) Top panel, differential
conductance $dI/dV$ plot as a function of $V_{sd}$ and $V_{g}$
at base temperature and zero magnetic field. The resonance features
are labeled as: $\circ$ for the main resonance of high conductance
and $\diamond$ for the weak conductance resonance. The letters e
and o represents the even and odd charge number states, and blue or
green arrows denote the ABSs in the charge state is $\pi$-type or
0-type, respectively. Bottom panel, zero-bias conductance $dI/dV$ over the same $V_{g}$ range.
(e)-(g) Traces of the $dI/dV$ taken at $V_{g}$ values indicated by the colored dashed lines in (d), respectively.
Inset of (g) is a current bias measurement at the same $V_{g}$, revealing transport with a supercurrent in the device.}
\label{fig.1}
\end{figure}
\newpage

\begin{figure}
\centering{}\includegraphics[width=0.83\textwidth]{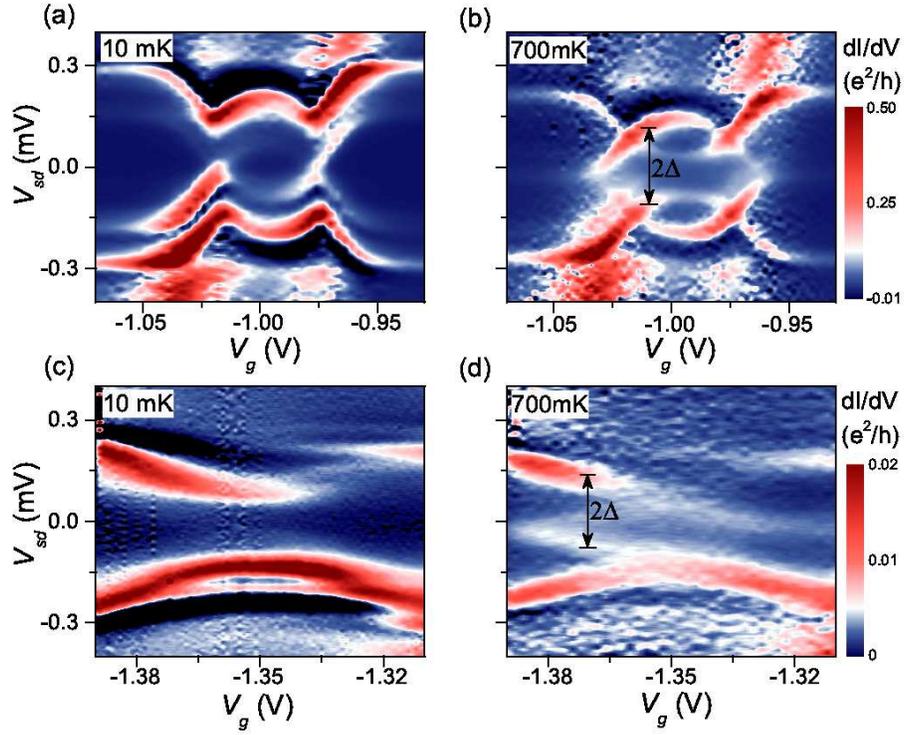} \caption{Temperature evolution  of different types of ABSs.(a) (b) $dI/dV(V_{sd},V_{g})$ for the odd charge
state at $V_{g}\simeq-1.0$ V measured at (a): base temperatures $T\sim10$ mK
and (b): $T\sim700$ mK. (c)
(d) $dI/dV(V_{sd},V_{g})$ for the odd charge
state at $V_{g}\simeq-1.35$ V measured at (c): base temperatures $T\sim10$ mK
and (d): $T\sim700$ mK. The distance labeling in (b) and (d) indicates the
energy spacing between the original ABS resonance and the thermally
developed ABS is a constant 2$\Delta$, i.e. they are parallel. }
\label{fig.2}
\end{figure}
\newpage

\begin{figure}
\centering{}\includegraphics[width=0.69\textwidth]{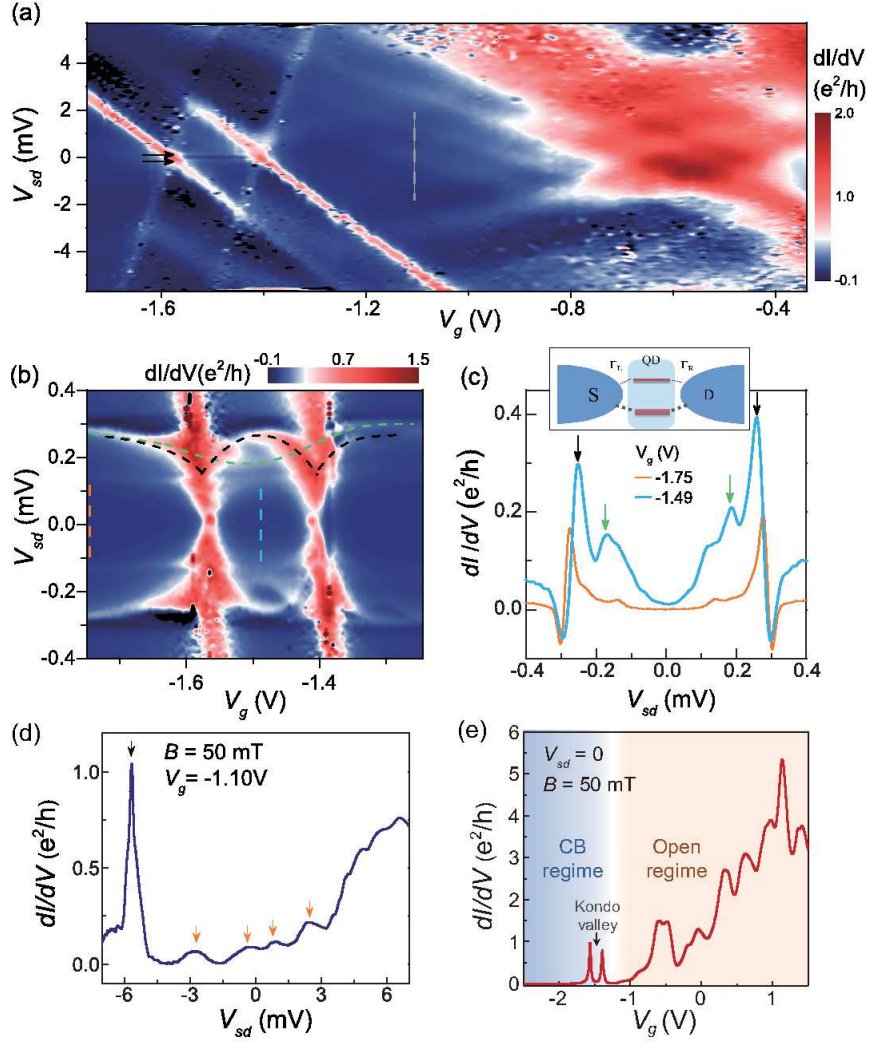} \caption{Characterization of the S-NWQD-S device in normal and superconducting state. (a) Normal-state differential conductance $dI/dV(V_{sd},V_{g})$ measured for a two terminal S-NWQD-S device (D2) with $L\sim65$ nm in contact separation
and $D\sim90$ nm in nanowire diameter at $T$ = 10 mK and $B$
= 50 mT. (b) Low bias differential conductance $dI/dV(V_{sd},V_{g})$ measured at $B=0$ of the charge state corresponding
to the diamond region around $V_{g} \sim -1.5$ V shown
in (a). The black and green dashed curves are guides for the two different types of
ABSs resonances. (c) $dI/dV(V_{sd})$ curves at selected $V_{g}$. The blue (orange) curve corresponds to the blue (orange) dashed line cut in panel (b). The black (green) arrows refer to the resonances denoting the $\pi(0)$-type ABSs. A
shoulder structure at around $|V_{sd}|\sim150 \ \upmu$eV originates from
the first-order multiple Andreev reflection. Inset: schematics of
the model of our S-NWQD-S system. The red bars represent the weakly
coupled QD level (thin) and the strongly coupled QD level (thick) which are
both involved in the transport processes.
(d) Normal state $dI/dV(V_{sd})$ trace taken at the grey dashed line cut in panel (a). The broad resonances at
low bias are indicated by orange arrows and the sharp resonance at high bias is indicated by the black arrow.
(e) Normal state $dI/dV(V_{g})$ trace for the S-NWQD-S device at $V_{sd}=0$. A quick transition of transport from a Coulomb blockade regime to an open regime can be recognized from the curve.}
\label{fig.3}
\end{figure}
\newpage

\begin{figure}
\centering{}\includegraphics[width=0.86\textwidth]{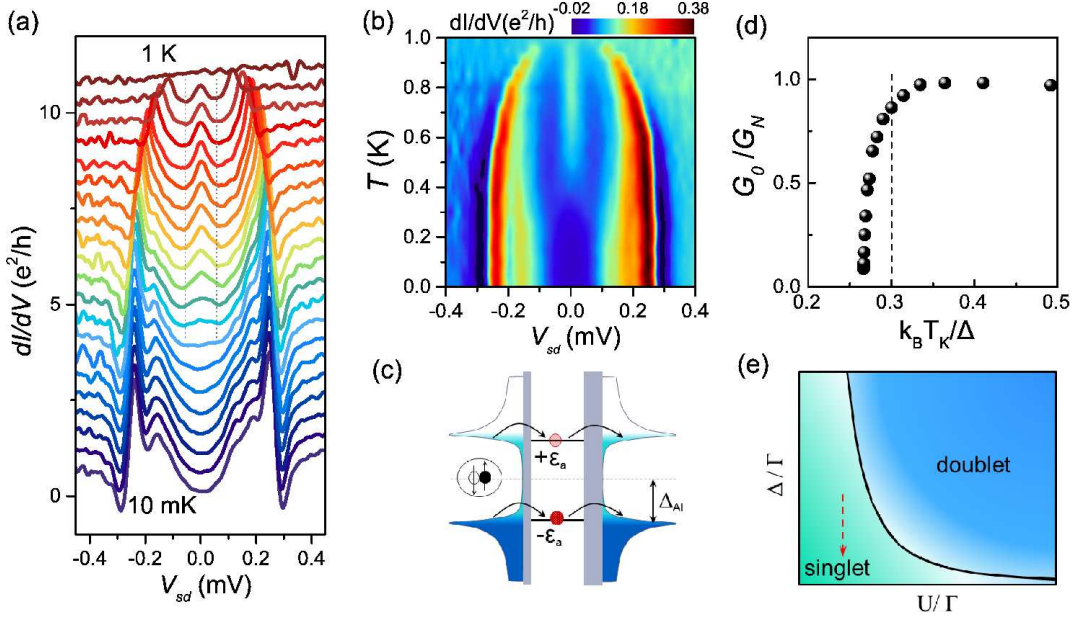} \caption{Temperature evolution of the differential conductance in the Kondo valley. (a) $dI/dV(V_{sd})$ curves measured
at increasing temperatures from 10 mK to 1 K at $B=0$, taken at $V_{g}= -1.49$ V denoted by the blue dashed
line cut in Fig. 3(b). A zero bias conductance peak can be identified
from above $T=400$ mK and vanishes at the critical temperature $T=1$ K. Curves are successively offset upward by 0.05$e^{2}/h$ for clarity. (b) 2D plot of  $dI/dV$ versus $V_{sd}$ and $T$. (c) Schematic of
the formation of the zero bias conductance peak resulting from the
ABS-assisted resonant tunneling of the thermally excited quasiparticles.
(d) Normalized zero-bias conductances, $G_{0}/G_{N}$, versus $T_{K}/\Delta$ obtained from traces in (a), where $G_{0}$ is the zero-bias conductance and $G_{N}$ is the normal state conductance at high bias.
(e) Schematic phase diagram for the singlet-doublet transition as a function of $\Delta/\Gamma$ and $U/\Gamma$.}
\label{fig.4}
\end{figure}

\newpage{}

\begin{figure}
\centering{}\includegraphics[width=0.86\textwidth]{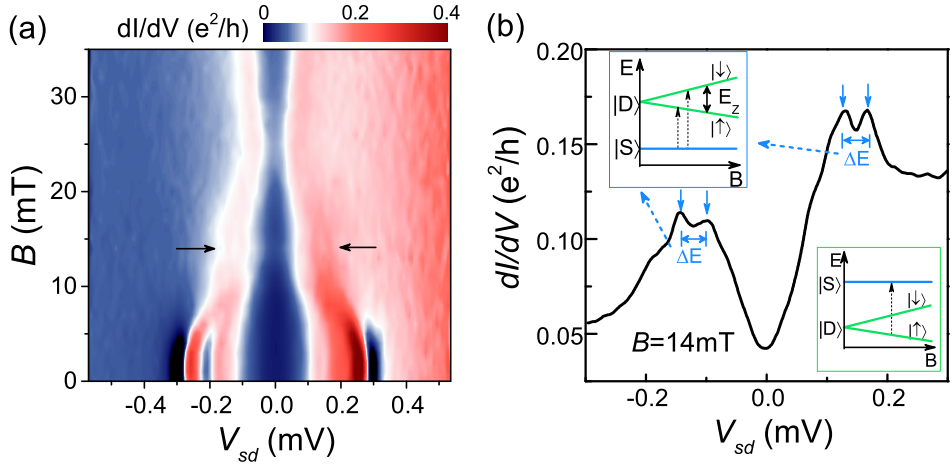} \caption{Magnetic field evolution of the differential conductance in the Kondo valley. (a) 2D plot of $dI/dV(V_{sd}, B)$
measured at $V_{g}=-1.49$ V denoted by the blue dashed line
cut in Fig. 3(b). A weak splitting in the 0-type ABS resonances is
observed at $B\sim14$ mT, indicated by a pair of black
arrows. (b) The $dI/dV(V_{sd})$ trace taken at $B=14$ mT. A split double-peak
can be seen on top of each 0-type ABS resonance. The spacing between
the double peaks is denoted by $\Delta E$ , which is expected to
be equal to the Zeeman energy $E_{Z}$. Upper left inset: schematic
of the splitting of the 0-type ABS. For a singlet ($|S\rangle$) ground
state and a doublet ($|D\rangle$) excited state, the
ABS will split under a magnetic field due to the Zeeman-splitting of
the doublet excited state. Lower right inset: schematic of the energy
change of the $\pi$-type ABS. Due to the Zeeman-splitting of doublet
ground state, the energy of the ABS will increase.}
\label{fig.5}
\end{figure}

\end{document}